\documentclass[preprint,epsf,aps,floats]{revtex4}
\usepackage{amsmath,amssymb,bm,epsfig,psfrag}
\usepackage{mathtools}

\usepackage{color}
\usepackage{float}

\newcommand{\be}{\begin{equation}}
\newcommand{\ee}{\end{equation}}
\newcommand{\ber}{\begin{eqnarray}}
\newcommand{\eer}{\end{eqnarray}}

\newcommand\ket[1]{|{#1}\rangle}

\def\Dsl{\,\raise.15ex \hbox{/}\mkern-12.8mu D}

\newcommand{\kk}{\textbf{k}}
\newcommand{\pp}{\textbf{p}}
\newcommand{\xx}{\textbf{x}}
\newcommand{\yy}{\textbf{y}}

\begin{document}

\title{Biexciton State Energies from Many-Body Perturbation Theory Based on Density Functional Theory Simulation}
\author{Deyan~Mihaylov,~Andrei~Kryjevski}
\affiliation{Department of Physics,~North Dakota State University,~Fargo, ND~58108,~USA}

\begin{abstract}
We develop a method for computing self-energy of a biexciton state in a semiconductor nanostructure 
using many-body perturbation theory (MBPT) based on the density functional theory (DFT) simulation.
We compute energies of low-energy biexciton states composed of singlet excitons in the chiral single-wall 
carbon nanotubes (SWCNT), such as (6,2), (6,5) and (10,5). In all cases we find a small decrease in the 
biexciton gap: -0.045 $eV$ in (6,2), which is 4.59\% of the non-interacting biexciton gap; -0.041 $eV$ in (6,5), 
which is 4.47\% of the non-interacting gap and -0.036 $eV$ in (10,5), which is 4.31\%.
\end{abstract}

\date{\today}

\maketitle

\section{Introduction}

\begin{figure}[h]
  \centering
  \includegraphics[width=0.4\textwidth]{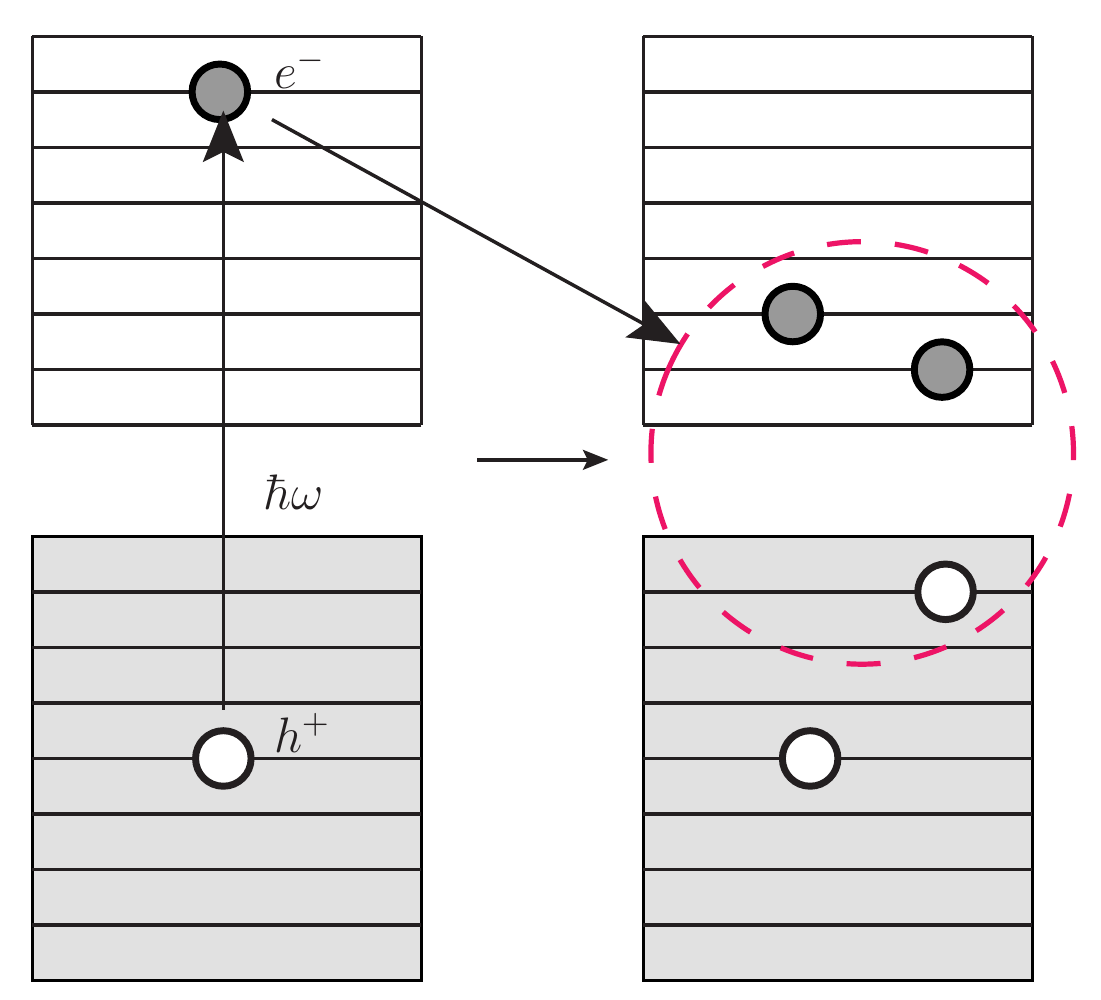}  
  \caption{Schematic illustration of the MEG process. Absorption of a photon with energy $\hbar \omega \ge 2E_{gap}$ creates and exciton which can decay into a biexciton.  
Pictured here, a hot electron loses some of its energy to the creation of another electron-hole pair. } 
\label{MEGprocess}   
\end{figure}

Efficient photon-to-electron energy conversion is an important property of a nanomaterial, which has been receiving a 
lot of attention. One mechanism of increasing the conversion efficiency is called multiple-exciton generation (MEG), where 
absorption of a high-energy photon results in the generation of multiple charge carriers. Fig. \ref{MEGprocess} illustrates 
the MEG process: absorption of a high-energy photon creates an electron-hole pair (exciton) with the energy exceeding twice 
the bandgap. In MEG, this excess photon energy is diverted into generation of additional charge carriers instead of being   
lost to generating atomic vibrations.  


It has been shown that biexciton bound states play an important role in the excited-state properties of SWCNTs and that exciton-exciton binding energies are strongly dependent on the nanotube chirality \cite{Matsunaga, Santos}. Experimental work aimed at addressing specifically the importance of exciton-exciton interactions in chiral SWCNTs has been performed by Colombier \textit{et al.}, where biexciton binding energy of 106 meV in (9,7) SWCNT has been reported \cite{Colombier}.  Previous theoretical work on exciton and biexciton binding energies in carbon nanotubes has been reported by Kammerlander \textit{et al.} where the quantum Monte Carlo method has been combined with the tight-binding approximation and binding energies of $\approx 120-150$ meV were reported \cite{Kammerlander}.
Theoretical efforts based on MBPT for calculating exciton-to-biexciton rates in MEG calculations in carbon nanotubes have been reported by Rabani \textit{et al.}, although in their work, the final biexciton state is treated as a pair of two non-interacting excitons \cite{Rabani}. {Additionally, there has been extensive theoretical work based on perturbation theory aimed at predicting carrier multiplication (CM) rates in semiconductor nanostructures \cite{Voros, Marri, Marri2}, but exciton-exciton interactions in the final biexciton state are not included. 
Explicit, first-principles treatment of exciton-exciton interactions in semiconductor nanostructures is reported by Piryatinski \textit{et al.} \cite{Piryatinski}, 
where first-order perturbation theory is applied to biexciton states in type II core/shell nanocrystals, where spatial separation of opposite charges and enhanced 
confinement of like-charges leads to large ($\sim$ 100 meV) positive shifts to the bi-exciton energies. However, the methodology developed in \cite{Piryatinski} is not 
applicable to biexciton states where exciton separation distance is comparable to electron-hole separation distance, as can be expected in CNTs.}   

Recently, Kryjevski \textit{et al.} have developed several methods for a comprehensive description of MEG in a nanosctructure using 
DFT-based MBPT, including exciton effects \cite{KryjevskiMEG2,KryjevskiMEG3,KryjevskiCNT,KryjevskiSF,KryjevskiMEG,KryjevskiCT}. 
First, one uses DFT-based MBPT technique to compute exciton-to-biexciton decay and recombination rates, {\it i.e.}, the rates of 
the inverse and direct Auger processes, respectively \cite{KryjevskiMEG2, KryjevskiMEG3}. Next, one utilizes DFT software to compute 
phonon frequencies and normal modes, 
which are then used to compute one- and two-phonon exciton emission rates \cite{KryjevskiMEG}. Finally, all these rates are incorporated 
into the Boltzmann transport equation (BE) which provides comprehensive nonperturbative description of time evolution of the excited state 
including ``competition" between different relaxation channels, such as MEG, phonon-mediated relaxation, {\it etc.} \cite{KryjevskiMEG}. 
In particular, one can compute number of excitons generated from a single high-energy photon, which is the internal quantum efficiency (QE). 
This method has been successfully applied to several chiral SWCNTs and efficient low-energy MEG was predicted, in a good agreement with the 
available experimental data \cite{MEGExperiment, MEGExperiment2}. Further, when augmented with the exciton transfer terms 
this BE technique has been applied to the doped SWCNT-$Si$ nanocrystal system and formation of a long-lived charge transfer state was predicted \cite{KryjevskiCT}. 
Additionally, in \cite{KryjevskiSF} MEG rates method for the singlet fission (SF) process, where a singlet exciton decays into two spin-one (triplet) excitons in 
the overall singlet state, has been developed and applied to SWCNTs.  

In all the MEG work so far 
the final biexciton state has been approximated as a non-interacting exciton pair. However, knowledge of the low-energy biexciton energy levels is essential 
for the accurate prediction of the MEG threshold. So, here we investigate this issue by developing a DFT-based MBPT method for biexciton state energies by including 
residual electrostatic (dipole-dipole) interactions between the excitons in the biexciton state. The technique is then applied to chiral SWCNTs (6,2), (6,5) and (10,5). Here, only biexciton states composed of two singlet excitons are included.

The paper is organized as follows. Section \ref{theory} contains description of the methods and
approximations employed in this work. Section \ref{CompDetail} contains description of the atomistic
models studied in this work and of DFT simulation details. Section \ref{Results} contains discussion
of the results obtained. Conclusions and Outlook are presented in Section \ref{Conclusion}. 

\section{Theoretical Methods and Approximations} \label{theory}

\subsection{Microscopic Hamiltonian} \label{Hamiltonian}

For completeness, let us review basics of the DFT-based MBPT approach that includes electron-exciton terms \cite{KryjevskiSF}. 

The electron field operator $\psi_{\sigma}(\xx)$ is related to the annihilation operator of the $i^{th}$ Kohn-Sham (KS) state $a_{i\alpha}$ via
\ber
  \psi_{\sigma}(\xx) = \sum_{i, \sigma} \phi_{i\sigma}(\xx)a_{i\sigma}
\eer 
where $\phi_{i\sigma}(\xx)$ is the $i^{th}$ KS orbital with spin $\sigma$ and $a_{i\sigma}, a_{i\sigma}^{\dagger} $ obey fermion anti-commutation relations $\{a_{i\sigma}, a_{j\nu}^{\dagger}\} = \delta_{ij}\delta_{\sigma\nu}, \{a_{i\sigma}, a_{j\nu}\} = 0$ \cite{Fetter,Mahan}. In the spin nonpolarized case we consider here, 
$\phi_{i\uparrow}(\xx)=\phi_{i\downarrow}(\xx)\equiv \phi_{i}(\xx).$ In terms of $a_{i\sigma}, a_{i\sigma}^{\dagger}$, the electron Hamiltonian is

\be
  {H}= \sum_{i\sigma} \varepsilon_i a_{i\sigma}^{\dagger} a_{i\sigma} + H_C - H_V + H_{e-exciton} \label{Hamiltonian} 
\ee 
where $\varepsilon_{i\uparrow} = \varepsilon_{i\downarrow} \equiv \varepsilon_{i}$ is the energy of the $i^{\text{th}}$ KS orbital. In a periodic structure KS energies and orbitals 
are labeled by the band number and lattice wavevector but here, as explained in Sec. \ref{CompDetail}, the state label is just an integer. The $H_C$ term is the microscopic Coulomb interaction operator  
\ber
  & H_C = \frac{1}{2} \sum_{ijkl \sigma, \nu} V_{ijkl} a^{\dagger}_{i \sigma} a^{\dagger}_{j \nu} a_{l \nu} a_{k \sigma} , \\
  & V_{ijkl}  = \int d\xx d\yy \phi_i^*(\xx)\phi_j^*(\yy) \frac{e^2}{| \xx - \yy |} \phi_l(\yy)\phi_k(\xx), 
\label{COULOMB}
\eer
where KS indices $i,j,k,l,...$ can refer to both occupied and unoccupied states within the included range. The $H_V$ term prevents double-counting of electron interactions 
\be
   H_V = \sum_{ij, \sigma} a^{\dagger}_{i \sigma} \bigg( \int d\xx d\yy \phi_i^*(\xx) V_{KS}(\xx, \yy) \phi_j(\yy) \bigg) a_{j \sigma} ,
\label{HV}
\ee
where $V_{KS}(\xx, \yy)$ is the Kohn-Sham potential used in the DFT simulation \cite{Onida,Kummel}.  

The electron-exciton coupling term $H_{e-exciton}$ appearing in Eq. \eqref{Hamiltonian} is
\ber
& H_{e-exciton} = \sum_{e h \alpha} \sum_{\sigma} \frac{1}{\sqrt{2}} \bigg( [\varepsilon_{e} - \varepsilon_h - E^{\alpha}]\Psi_{eh}^{\alpha} a_{h \sigma} a^{\dagger}_{e \sigma}({\rm B}^{\alpha} + {\rm B}^{\alpha \dagger}) + h.c. \bigg) + \sum_{\alpha} E^{\alpha} {\rm B}^{\alpha \dagger} {\rm B}^{\alpha}  \label{EXCFERMSF} 
\eer
where $e \ge$ LU (the lowest unoccupied KS state) and $h \le HO$ (the highest occupied KS state), ${\rm B}^{\alpha \dagger}$, $\Psi_{eh}^{\alpha}$ and  $E^{\alpha}$ 
are the singlet exciton creation operator, wave function and energies respectively. Technically, the full electron Hamiltonian is comprised of only the first three 
terms in Eq. \eqref{Hamiltonian}, however adding $H_{e-exciton}$ along with the rules to avoid double counting allows for the perturbative treatment of excitonic 
effects and is a standard approch to describing the coupling between excitons and electrons and holes \cite{Beane, Spataru1, Perebeinos, Spataru2}. The $H_{e-exciton}$ term is a result of resummation of Coulomb perturbative corrections resulting from the Coulomb interaction operator $H_C$ (Eq. \eqref{COULOMB}) to the electron-hole state, 
which is an implementation of a standard method to include bound states into the MBPT framework \cite{Berestetskii, Beane, KryjevskiSF}.      

Exciton wave functions, $\Psi_{eh}^{\alpha},$ and energies, $E^{\alpha},$ are solutions to the Bethe-Salpeter equation (BSE), which is (see, {\it e.g.}, \cite{Benedict})       
\be
   [\varepsilon_{e} - \varepsilon_h - E^{\alpha}]\Psi_{eh}^{\alpha} + \sum_{e' h'} (K_{C} + K_{D})(e,h;e',h')\Psi_{e'h'}^{\alpha} = 0, \label{BSE}
\ee
where
\be
  K_{C} =\frac{8 \pi e^2}{V} \sum_{\kk \neq 0} \frac{\rho_{eh}(\kk)\rho_{e'h'}^*(\kk)}{k^2}, \label{KCOUL}
\ee
\be
  K_{D} = - \frac{4 \pi e^2}{V} \sum_{\kk \neq 0} \frac{\rho_{ee'}(\kk)\rho_{hh'}^*(\kk)}{|\kk|^2 - \Pi(0, -\kk, \kk)}, \label{KDIR}
\ee
where
\be
  \rho_{ij}(\kk) = \sum_{\pp} \phi_j^*(\pp - \kk) \phi_i(\pp) 
\ee

{Medium screening is taken into account via the polarization function (Eq. \eqref{PI}) which appears only in the direct term (Eq. \eqref{KDIR}) of BSE.} In the random-phase approximation (RPA) 
\be
   \Pi(\omega,\kk,\pp) = \frac{8 \pi e^2}{\hbar V} \sum_{ij} \rho_{ij}(\kk)\rho_{ji}(\pp)\bigg(\frac{\theta_{-j}\theta_{i}}{\omega - \omega_{ij} + i\delta} - \frac{\theta_{j}\theta_{-i}}{\omega - \omega_{ij} - i\delta} \bigg), \;\;\; \omega_{ij} = \frac{\varepsilon_i - \varepsilon_j}{\hbar} \label{PI}
\ee
where $V$ is the simulation cell volume, $\delta$ is the width parameter, which will be set to $0.025~eV$ corresponding to the room temperature scale.  

Here, we use static approximation, taking  $\Pi(\omega=0,\kk,\pp).$ This is a widely-used approach for semiconductor nanostructures ({\it e.g.}, \cite{PhysRevLett.90.127401,PhysRevB.68.085310,PhysRevB.79.245106}), which is justified by the cancellations that appear when the electron-hole screening and the single-particle Green's functions are  
both treated dynamically \cite{Bechstedt}. Next, the main simplifying approximation employed in this work is to retain only the diagonal elements of the polarization matrix, 
\textit{i.e.}, $\Pi(0,\kk,\pp) \simeq \Pi(0,\kk,-\kk)\delta_{\kk, -\pp} $, or $\Pi(0,\xx,\xx') \simeq \Pi(0,\xx - \xx')$ in position space, {\it i.e.}, the system 
is approximated as a uniform medium. This is a valid approximation for quasi-one-dimensional systems, such as CNTs,  where one can expect $\Pi(\xx,\xx') \simeq \Pi(z-z')$, with $z,z'$ being the axial positions. This diagonal polarization matrix approximation, which is a an improvement on previous studies on CNTs in which screening has been approximated by a dielectric constant \cite{Perebeinos2}, has been employed for the time being, as it significantly reduces computational costs. Calculations including full treatment of the polarization matrix, although unlikely to change qualitative conclusions, are left for future work. Also, in the DFT simulations one uses hybrid Heyd-Scuseria-Ernzerhof (HSE06) exchange-correlation functional \cite{HSE06}   
which is to substitute for the $G_0W_0$ calculation of single-particle energies - the second step in the standard three-step process in the electronic structure  
calculation \cite{Rohlfing2,Hybertsen}. The HSE06 functional, which is significantly less computationally expensive than $G_0W_0$, has been shown to produce somewhat  reasonable 
results for bandgaps in semiconducting nanostructures \cite{Kummel, HSE06Muscat, LouieHSE06}. So, single-particle energies and orbitals are approximated by the KS $\varepsilon_i$ and 
$\phi_i({\bf x})$ from the HSE06 DFT output. Therefore, in our approximation using HSE06 replaces ``dressing'' fermion 
lines in the Feynman diagrams, including subtraction of the compensating term  (\ref{HV}). 

For SWCNTs, the set of approximations stated above have been checked and shown to be reasonable by reproducing experimental results for the low-energy absorption peaks in (6,2), (6,5) and 
(10,5) SWCNTs within 5-13\% error \cite{KryjevskiMEG}.
     

\section{Expressions for the Biexciton Self-Energy} \label{Expressions}

In order to account for the exciton-exciton interactions one computes self-energy of the biexciton state $\hbar\Sigma_{\alpha\beta} (E)$. 
Then the biexciton energy is approximated as   
\ber
 E_{\alpha\beta}=E^{\alpha} + {E}^{\beta}+\hbar\Sigma_{\alpha\beta} (E = E^{\alpha} + E^{\beta}) \label{self_energy_corr}
\eer
where $E^{\alpha}$ and $E^{\beta}$ are the energies of the two excitons in the biexciton state. $E^{\alpha}$ and $E^{\beta}$ are obtained by solving the Bethe-Salpeter equation (Eq. \eqref{BSE}). $\hbar\Sigma_{\alpha\beta} (E = E^{\alpha} + E^{\beta})$ is the biexciton self-energy evaluated on the total energy of a state composed of two non-interacting excitons $\ket{\alpha}$ and $\ket{\beta}$. 

\begin{figure}[H]
  \centering
  \includegraphics[width=0.97\textwidth]{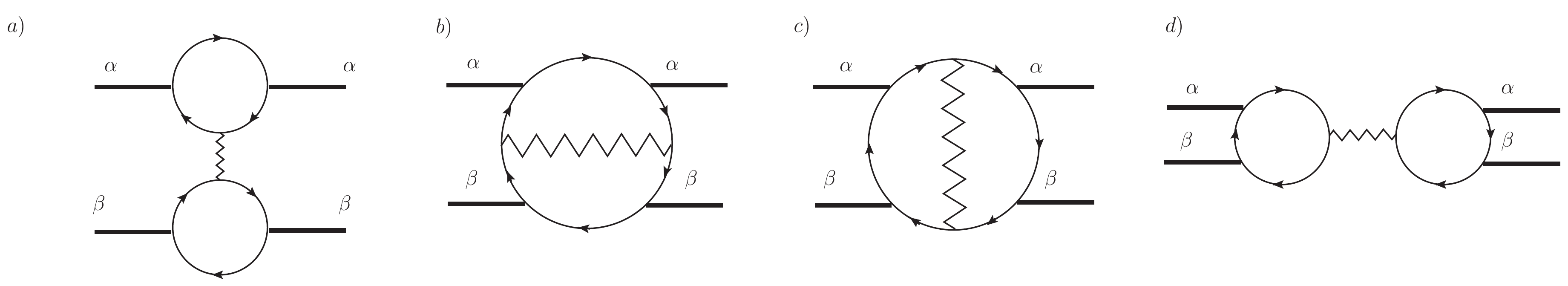}
  \caption{Leading order Feynman diagrams representing contributions to the biexciton self-energy. Thin (fermion) lines are Kohn-Sham particle/hole propagators, thick lines - excitons (see Fig. \ref{EXCbound}), zigzag lines - the {screened} Coulomb interaction. Not shown for brevity are similar diagrams where arrow directions in one of the fermion loops are reversed, and diagrams where all arrows are reversed.  
 } \label{BIEXCfig}
\end{figure}

To the leading order in the Coulomb interaction the relevant Feynman diagrams are shown in Fig. \ref{BIEXCfig}. One only retains 
corrections to the bare biexciton state where a particle or a hole from the exciton $\ket{\alpha}$ interacts with a particle or a hole 
from the other exciton $\ket{\beta}$; the interactions between particles and holes within the same exction are already included by the BSE. 
Note that including contributions with all possible arrow directions in each fermion loop is needed to include
all the interactions between electrons and holes within the biexciton state.

\noindent
Note that for each diagram in Fig. \ref{BIEXCfig} there is an additional contribution upon exchange of the $\alpha$ and $\beta$ excitons on the right-hand-side of the fermion loop. These contributions are currently a subject of future work. The expressions resulting from each diagram in Fig. \ref{BIEXCfig} are similar in general form, so for brevity, only one of them will be quoted here and the rest will be presented in the appendix. For instance, the diagram from Fig. \ref{BIEXCfig} - \textit{a)} results in the following expression
\ber
 & \hbar\Sigma_{\alpha\beta} (E = E^{\alpha} + E^{\beta}) = \frac{4}{5} \sum_{ijklmn} \frac{\Theta_{i}\Theta_n\Theta_{l}\Theta_{-k}\Theta_{-j}\Theta_{-m}}{(E^{\alpha}-\epsilon_{nk} - i\delta)(E^{\beta}-\epsilon_{lm}+ i\delta)(\epsilon_{kj} - i\delta)(\epsilon_{li} + i\delta)} \times  \nonumber  \\ 
& \times (\Psi_{im}^{\beta})^*(E^{\beta} - \epsilon_{im})(\Psi_{nk}^{\alpha})^*(E^{\alpha} - \epsilon_{nk}) \Psi_{lm}^{\beta} (E^{\beta} - \epsilon_{lm})  \Psi_{nj}^{\alpha} (E^{\alpha} - \epsilon_{nj}) \times W_{ijkl} \label{expression}
\eer
where $\Psi_{ij}^{\alpha}$ and $E^{\alpha}$ are the exciton wave function and exciton energies respectively obtained by solving the Bethe-Salpeter equation (Eq. \eqref{BSE}). $\epsilon_{ij} = \epsilon_i - \epsilon_j$; $W_{ijkl}$ are the RPA-screened Coulomb matrix elements  
\be
  W_{ijkl} = \frac{4 \pi e^2}{V} \sum_{\kk \neq 0} \frac{\rho_{il}^*(\kk)\rho_{jk}(\kk)}{|\kk|^2 - \Pi(0, -\kk, \kk)} \label{W}
\ee
where $\Pi(0, -\kk, \kk)$ is defined in Eq. \eqref{PI}. The theta-functions determine whether the KS indices $i,j,k,l$ are particles or holes
\be
\Theta_i = \sum_{i > \text{HO} } ~~ , ~~ \Theta_{-i} = \sum_{i \le \text{HO}}.
\ee
 In this work we use 
\ber
\frac{1}{x \pm i \delta}=\frac{x}{x^2+\delta^2} \mp i \frac{\delta}{x^2+\delta^2},
\label{1xidelta}
\eer
i.e., both the principal value and delta function parts of $1/(x \pm i \delta)$ factors are included.

The other three diagrams - \ref{BIEXCfig}-$b),~c),~d)$ - produce similar expressions not shown for the sake of brevity.  

\begin{figure}[H]
  \centering
  \includegraphics[width=0.89\textwidth]{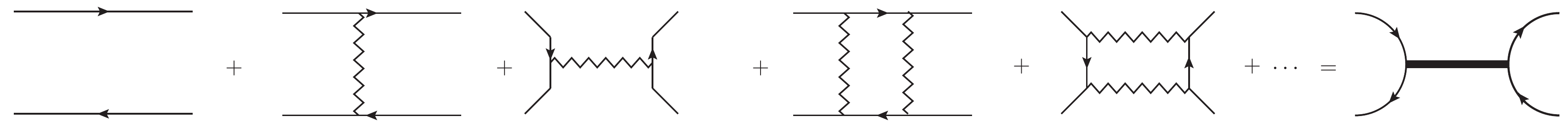}
  \caption{ {Summing perturbative contributions from Coulomb interactions (zigzag lines) between electron and hole (thin lines) is equivalent to including intermediate exciton bound state (thick line).} } \label{EXCbound}
\end{figure}

\begin{figure}[H]
  \centering
  \includegraphics[width=0.99\textwidth]{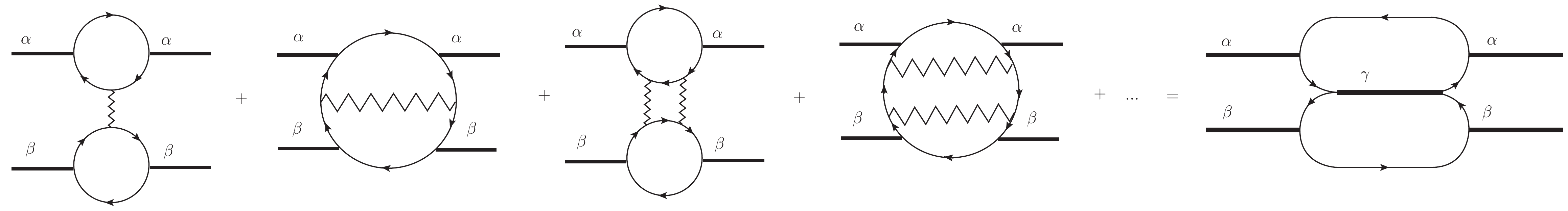}
  \includegraphics[width=0.99\textwidth]{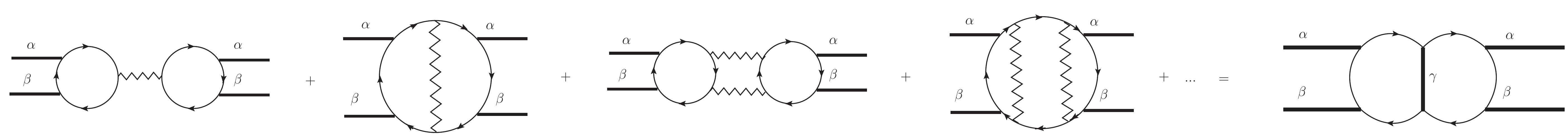}
  \caption{ Summing perturbative contributions from Coulomb interactions (zigzag lines) between electron and hole (thin lines) from different excitons  - $\alpha$ and $\beta$ - results in appearance of the intermediate exciton $\gamma$.} \label{BIEXCresum} 
\end{figure}

Next, the possibility of including higher order perturbative corrections was explored. Naively including terms of second (or higher) order in the Coulomb interaction, i.e., 
decorating diagrams in Fig. \ref{BIEXCfig} with additional zigzag lines, would lead to prohibitively expensive calculations, even for a small system. However, certain classes 
of perturbative contributions can be summed to all orders, {such as the Coulomb interactions between electron and hole resulting in formation of an exciton bound state (see Fig. \ref{EXCbound}). 
Eq. \ref{EXCFERMSF} describes the resulting exciton-electron coupling term. } 
\noindent
This summation of Coulomb corrections to all orders can be applied to modify the diagrams shown in Fig. \ref{BIEXCfig}. In Fig. \ref{BIEXCresum} it is illustrated how an intermediate exciton state 
$\gamma$ appears from decorating the electron-hole lines in Fig. \ref{BIEXCfig} with zigzag lines in all possible ways. Here, we work to the leading order in the electron-exciton coupling - Eq. \ref{EXCFERMSF} - and, so, only decorate one pair of the electron-hole lines.  

\begin{figure}[H]
  \centering
  \includegraphics[width=0.99\textwidth]{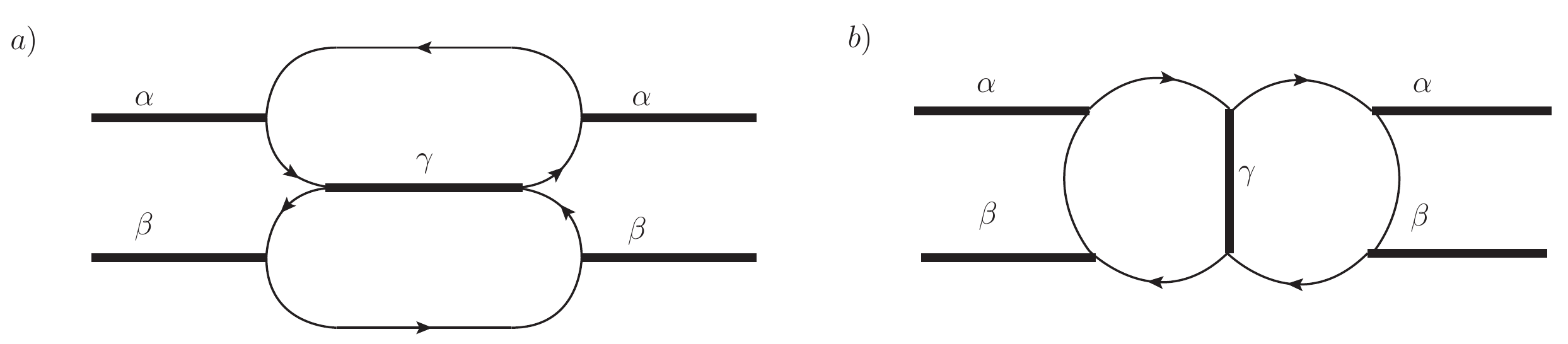}
  \caption{ \textit{a)} - diagram resulting from modification of diagrams \ref{BIEXCfig}-\textit{a)} and \textit{b)}.  \textit{b)} - diagram resulting from modification of diagrams \ref{BIEXCfig}-\textit{c)} and \textit{d)}.  } \label{BIEXCfig2}
\end{figure}

\noindent
The two distinct diagrams resulting from modifying Fig. \ref{BIEXCfig} are shown in Fig. \ref{BIEXCfig2}. 
\noindent
The contributions to  the biexciton self-energy 
from the diagrams in Fig. \ref{BIEXCfig2} are 
\ber
 & \hbar\Sigma_{\alpha\beta}^+ = \hbar\Sigma_{\alpha\beta}^{\textrm{a}} + \hbar\Sigma_{\alpha\beta}^{\textrm{b}} \label{expression+} \\
 & \hbar\Sigma_{\alpha\beta}^{\textrm{a}}(E = E^{\alpha} + E^{\beta})  = \frac{8}{45} \sum_{ijklmn\gamma} \frac{\Theta_{j}\Theta_k\Theta_{n}\Theta_{-i}\Theta_{-l}\Theta_{-m}}{(E^{\alpha} - \epsilon_{ki} + i\delta)(E^{\beta} - \epsilon_{nm}- i\delta)(E^{\gamma} - \epsilon_{km} + i\delta)(E^{\gamma} - \epsilon_{jl} + i\delta)(\epsilon_{jk} + i\delta)} \times  \label{expression2} \\  
& \times (\Psi_{ji}^{\alpha})^*(E^{\alpha} - \epsilon_{ji})\Psi_{ki}^{\alpha}(E^{\alpha} - \epsilon_{ki}) \Psi_{nl}^{\beta} (E^{\beta} - \epsilon_{nl})  (\Psi_{nm}^{\beta})^* (E^{\beta} - \epsilon_{nm}) \Psi_{jl}^{\gamma} (E^{\gamma} - \epsilon_{jl})  (\Psi_{km}^{\gamma})^* (E^{\gamma} - \epsilon_{km}) \nonumber  
\eer
\ber
 & \hbar\Sigma_{\alpha\beta}^{\textrm{b}}(E^{\alpha} + E^{\beta})  = \frac{8}{45} \sum_{ijklmn\gamma} \frac{\Theta_{j}\Theta_k\Theta_{n}\Theta_{-i}\Theta_{-l}\Theta_{-m}}{(E^{\alpha} - \epsilon_{kl} + i\delta)(E^{\beta} - \epsilon_{km}- i\delta)(E^{\gamma} - \epsilon_{jl} + i\delta)(E^{\gamma} - \epsilon_{nm} + i\delta)(\epsilon_{jn} + i\delta)} \times \label{expression3} \\ 
& \times (\Psi_{ji}^{\alpha})^*(E^{\alpha} - \epsilon_{ji})\Psi_{kl}^{\alpha}(E^{\alpha} - \epsilon_{kl}) \Psi_{ni}^{\beta} (E^{\beta} - \epsilon_{ni})  (\Psi_{km}^{\beta})^* (E^{\beta} - \epsilon_{km}) \Psi_{jl}^{\gamma} (E^{\gamma} - \epsilon_{jl})  (\Psi_{nm}^{\gamma})^* (E^{\gamma} - \epsilon_{nm}) \nonumber 
\eer
\noindent
where the additional summation label $\gamma$ is over the intermediate exciton state. Note that here the screened Coulomb interaction only appears implicitly through the exciton wavefunctions and energies. 

Additionally, there are self-energy contributions from the repulsive electron-electron and hole-hole interactions. These come from, e.g., diagrams in Fig. \ref{BIEXCfig} a) and d)
but with arrows in one of the fermion loops reversed. These corrections are treated to the leading order in the Coulomb interaction. There are eight distinct  expressions contributing to the biexciton self-energy due to the inter-exciton electron-electron or hole-hole interactions, two from each of the 4 diagrams in Fig. \ref{BIEXCfig} with the appropriate fermion arrow flips. The expressions are 
\ber
 & \hbar\Sigma_{\alpha\beta}^- = \hbar\Sigma_{\alpha\beta}^{\textrm{ee},a} + \hbar\Sigma_{\alpha\beta}^{\textrm{hh},a}+\hbar\Sigma_{\alpha\beta}^{\textrm{ee},b} + \hbar\Sigma_{\alpha\beta}^{\textrm{hh},b}+\hbar\Sigma_{\alpha\beta}^{\textrm{ee},c} + \hbar\Sigma_{\alpha\beta}^{\textrm{hh},c} +\hbar\Sigma_{\alpha\beta}^{\textrm{ee},d} + \hbar\Sigma_{\alpha\beta}^{\textrm{hh},d} \label{expression-}\\         
 & \hbar\Sigma_{\alpha\beta}^{\textrm{ee},a} (E^{\alpha} + E^{\beta}) = -\frac{4}{5} \sum_{ijklmn} \frac{\Theta_{i}\Theta_{l}\Theta_{k}\Theta_{j}\Theta_{-m}\Theta_{-n}}{(E^{\alpha}+\epsilon_{kn} - i\delta)(E^{\beta}-\epsilon_{lm}+ i\delta)(\epsilon_{kj} - i\delta)(\epsilon_{li} + i\delta)} \times  \nonumber  \\ 
& \times (\Psi_{im}^{\beta})^*(E^{\beta} - \epsilon_{im})(\Psi_{kn}^{\alpha})^*(E^{\alpha} - \epsilon_{kn}) \Psi_{lm}^{\beta} (E^{\beta} - \epsilon_{lm})  \Psi_{jn}^{\alpha} (E^{\alpha} - \epsilon_{jn}) \times W_{ijkl} \label{expression4} \\
& \hbar\Sigma_{\alpha\beta}^{\textrm{hh},a} (E^{\alpha} + E^{\beta}) = -\frac{4}{5} \sum_{ijklmn} \frac{\Theta_{-i}\Theta_{-l}\Theta_{-k}\Theta_{-j}\Theta_{m}\Theta_{n}}{(E^{\alpha}+\epsilon_{nk} - i\delta)(E^{\beta}-\epsilon_{ml} + i\delta)(\epsilon_{kj} - i\delta)(\epsilon_{li} + i\delta)} \times  \nonumber  \\ 
& \times (\Psi_{mi}^{\beta})^*(E^{\beta} - \epsilon_{mi})(\Psi_{nk}^{\alpha})^*(E^{\alpha} - \epsilon_{nk}) \Psi_{ml}^{\beta} (E^{\beta} - \epsilon_{ml})  \Psi_{nj}^{\alpha} (E^{\alpha} - \epsilon_{nj}) \times W_{ijkl} \label{expression5} \\
& \hbar\Sigma_{\alpha\beta}^{\textrm{ee},b} (E^{\alpha} + E^{\beta}) = -\frac{4}{5} \sum_{ijklmn} \frac{\Theta_{i}\Theta_{l}\Theta_{k}\Theta_{j}\Theta_{-m}\Theta_{-n}}{(E^{\alpha}+\epsilon_{im} - i\delta)(E^{\beta}-\epsilon_{ln}+ i\delta)(\epsilon_{il} - i\delta)(\epsilon_{jk} + i\delta)} \times  \nonumber \\
& \times (\Psi_{kn}^{\beta})^*(E^{\beta} - \epsilon_{kn})(\Psi_{jm}^{\alpha})^*(E^{\alpha} - \epsilon_{jm}) \Psi_{ln}^{\beta} (E^{\beta} - \epsilon_{ln})  \Psi_{im}^{\alpha} (E^{\alpha} - \epsilon_{im}) \times W_{ijkl} \label{expression6} \\
& \hbar\Sigma_{\alpha\beta}^{\textrm{hh},b} (E^{\alpha} + E^{\beta}) = -\frac{4}{5} \sum_{ijklmn} \frac{\Theta_{-i}\Theta_{-l}\Theta_{-k}\Theta_{-j}\Theta_{m}\Theta_{n}}{(E^{\alpha}-\epsilon_{mi} - i\delta)(E^{\beta}-\epsilon_{nl} + i\delta)(\epsilon_{il} - i\delta)(\epsilon_{jk} + i\delta)} \times  \nonumber  \\ 
& \times (\Psi_{nk}^{\beta})^*(E^{\beta} - \epsilon_{nk})(\Psi_{mj}^{\alpha})^*(E^{\alpha} - \epsilon_{mj}) \Psi_{nl}^{\beta} (E^{\beta} - \epsilon_{nl})  \Psi_{mi}^{\alpha} (E^{\alpha} - \epsilon_{mi}) \times W_{ijkl} \label{expression7} \\
& \hbar\Sigma_{\alpha\beta}^{\textrm{ee},c} (E^{\alpha} + E^{\beta}) = -\frac{4}{5} \sum_{ijklmn} \frac{\Theta_{i}\Theta_{l}\Theta_{k}\Theta_{j}\Theta_{-m}\Theta_{-n}}{(E^{\alpha}+\epsilon_{im} - i\delta)(E^{\beta}-\epsilon_{jm}+ i\delta)(\epsilon_{il} - i\delta)(\epsilon_{jk} + i\delta)} \times  \nonumber \\
& \times (\Psi_{kn}^{\beta})^*(E^{\beta} - \epsilon_{kn})(\Psi_{ln}^{\alpha})^*(E^{\alpha} - \epsilon_{ln}) \Psi_{jm}^{\beta} (E^{\beta} - \epsilon_{jm})  \Psi_{im}^{\alpha} (E^{\alpha} - \epsilon_{im}) \times W_{ijkl} \label{expression8} \\
& \hbar\Sigma_{\alpha\beta}^{\textrm{hh},c} (E^{\alpha} + E^{\beta}) = -\frac{4}{5} \sum_{ijklmn} \frac{\Theta_{-i}\Theta_{-l}\Theta_{-k}\Theta_{-j}\Theta_{m}\Theta_{n}}{(E^{\alpha}-\epsilon_{mi} - i\delta)(E^{\beta}+\epsilon_{mj}+ i\delta)(\epsilon_{il} - i\delta)(\epsilon_{jk} + i\delta)} \times  \nonumber \\
& \times (\Psi_{nk}^{\beta})^*(E^{\beta} - \epsilon_{nk})(\Psi_{nl}^{\alpha})^*(E^{\alpha} - \epsilon_{nl}) \Psi_{mj}^{\beta} (E^{\beta} - \epsilon_{mj})  \Psi_{mi}^{\alpha} (E^{\alpha} - \epsilon_{mi}) \times W_{ijkl} \label{expression9} \\
& \hbar\Sigma_{\alpha\beta}^{\textrm{ee},d} (E^{\alpha} + E^{\beta}) = -\frac{4}{5} \sum_{ijklmn} \frac{\Theta_{i}\Theta_{l}\Theta_{k}\Theta_{j}\Theta_{-m}\Theta_{-n}}{(E^{\alpha}-\epsilon_{im} - i\delta)(E^{\beta}+\epsilon_{lm}+ i\delta)(\epsilon_{il} + i\delta)(\epsilon_{kj} - i\delta)} \times  \nonumber \\
& \times (\Psi_{kn}^{\beta})^*(E^{\beta} - \epsilon_{kn})(\Psi_{jn}^{\alpha})^*(E^{\alpha} - \epsilon_{jn}) \Psi_{lm}^{\beta} (E^{\beta} - \epsilon_{lm})  \Psi_{im}^{\alpha} (E^{\alpha} - \epsilon_{im}) \times W_{ijkl} \label{expression10} \\
& \hbar\Sigma_{\alpha\beta}^{\textrm{hh},d} (E^{\alpha} + E^{\beta}) = -\frac{4}{5} \sum_{ijklmn} \frac{\Theta_{-i}\Theta_{-l}\Theta_{-k}\Theta_{-j}\Theta_{m}\Theta_{n}}{(E^{\alpha}+\epsilon_{mi} - i\delta)(E^{\beta}-\epsilon_{ml}+ i\delta)(\epsilon_{il} + i\delta)(\epsilon_{kj} - i\delta)} \times  \nonumber \\
& \times (\Psi_{nk}^{\beta})^*(E^{\beta} - \epsilon_{nk})(\Psi_{nj}^{\alpha})^*(E^{\alpha} - \epsilon_{nj}) \Psi_{ml}^{\beta} (E^{\beta} - \epsilon_{ml})  \Psi_{mi}^{\alpha} (E^{\alpha} - \epsilon_{mi}) \times W_{ijkl}, \label{expression11}
\eer
{ where, for example, $\hbar\Sigma_{\alpha\beta}^{\textrm{ee},a}$ stands for a self-energy contribution from a diagram representing electron-electron interactions obtained by flipping the arrows in one of the fermion loops in Fig. \ref{BIEXCfig}-\textit{a)}. Other symbols hold the same meaning as in Eq. \eqref{expression}. Note that here, Coulomb matrix elements $W_{ijkl}$ have either all-electron or all-hole indices. Thus, the complete expression for the biexciton self-energy is
\ber
\hbar\Sigma_{\alpha\beta} = \hbar\Sigma_{\alpha\beta}^+ + \hbar\Sigma_{\alpha\beta}^-
\eer
where $\hbar\Sigma_{\alpha\beta}^+$ is defined in \eqref{expression+} and $\hbar\Sigma_{\alpha\beta}^-$ is defined in \eqref{expression-}. }

\section{Atomistic Models}\label{CompDetail}

\begin{figure}[H]
  \centering
  \includegraphics[width=0.5\textwidth]{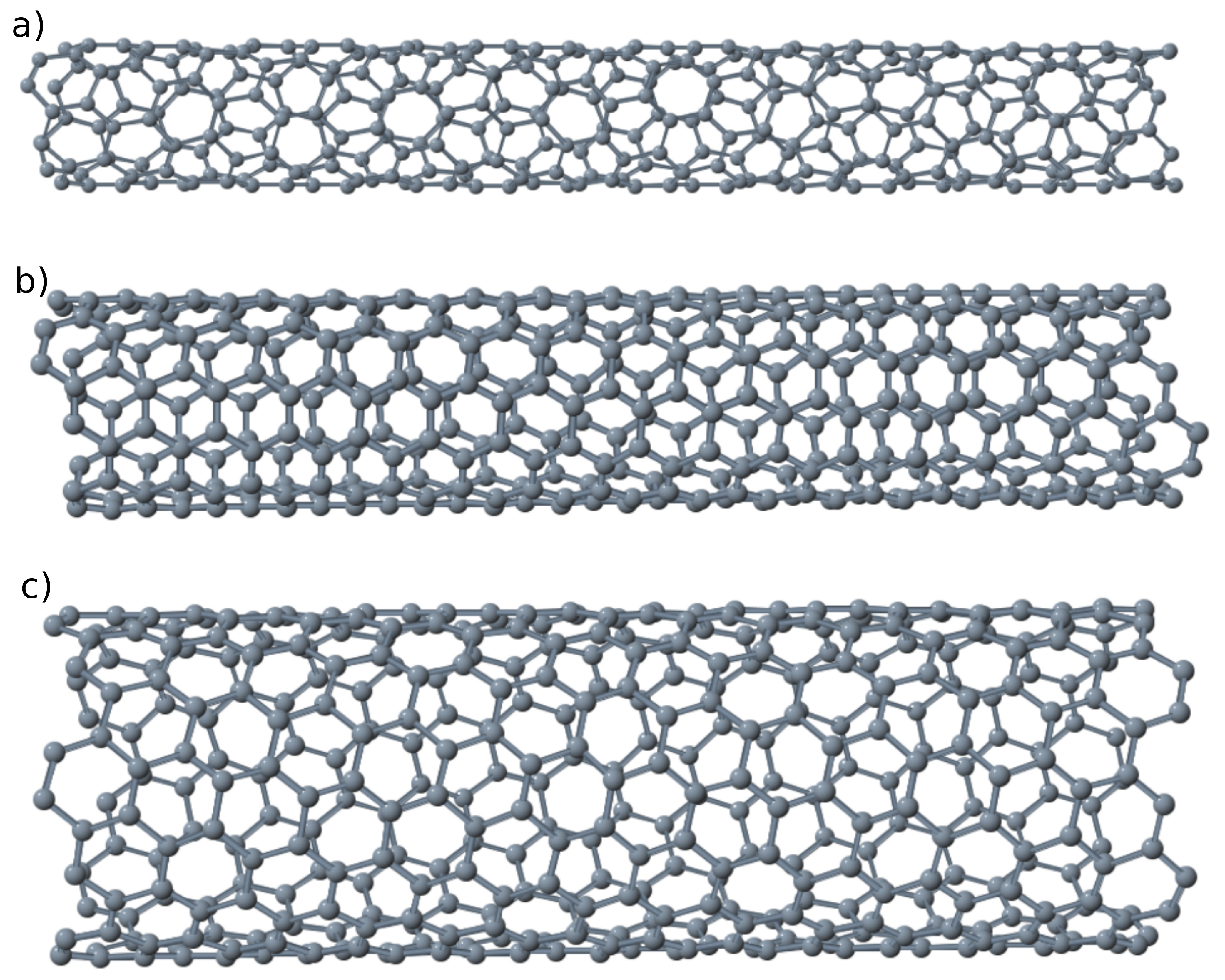} 
  \caption{ Atomistic models of the three chiral SWCNTs: a) - (6,2), b) - (6,5) and c) - (10,5)  } \label{CNTall} 
\end{figure}

Calculations of biexciton self-energies have been performed for chiral SWCNTs (6,5), (6,2) and (10,5), which are shown in Fig. \ref{CNTall}. DFT with HSE06 functional, as implemented in VASP (Vienna \textit{ab-initio} simulation package) \cite{VASP}, was used to optimize geometries and obtain KS orbitals $\phi_{i\sigma}(\xx)$ and energies $\varepsilon_i$. The momentum cutoff is defined as
\be
   \frac{\hbar^2 \kk^2}{2m} \le E_{max} , \;\;\;
   \kk = 2\pi \bigg( \frac{n_x}{L_x}, \frac{n_y}{L_y}, \frac{n_z}{L_z} \bigg), \;\;\; n_x, n_y, n_z = 0, \pm 1, \pm 2, \cdots  \label{cutoff}
\ee
where $m$ is the electron mass; we used $E_{max} = 300~ eV$. The number of orbitals used in the calculations was determined by the condition   
$E_{i_{max}} - E_{HO} \simeq E_{LU} - E_{i_{min}} \ge 3.5~eV$ where $i_{max}$/$i_{min}$ label the highest/lowest orbital. 

Periodic boundary conditions were used in the DFT simulations. In the axial direction the simulation cell length was chosen to accommodate 
an integer number of unit cells, while in the other two directions the SWCNT surfaces were separated by about 1 $nm$ of vacuum in order to avoid spurious interactions 
between their periodic images. For (6,2) and (10,5) the simulations have been performed including three unit cells. The rationale for this is that 
future work will involve SWCNTs with functionalized surfaces. So, including several unit cells will allow to keep the dopant concentration reasonably low. 
Also, including three unit cells instead of one substitutes for the Brillouin zone sampling. So, here we perform calculations at the $\Gamma$ point only. 
Previously, it has been shown that the variations in the single particle energies over the Brillouin zone is reasonably small ($\approx 10\%$) when three 
unit cells have been included in the simulation instead of one \cite{KryjevskiMEG}. For the (6,5) SWCNT only one unit cell was included due to high computational cost. 
However, the absorption spectrum for (6,5) was reproduced with the same accuracy as for the other two CNTs \cite{KryjevskiMEG}. 

\section{Results and Discussion} \label{Results} 
\begin{figure}[H]
  \centering
  \includegraphics[width=0.8\textwidth]{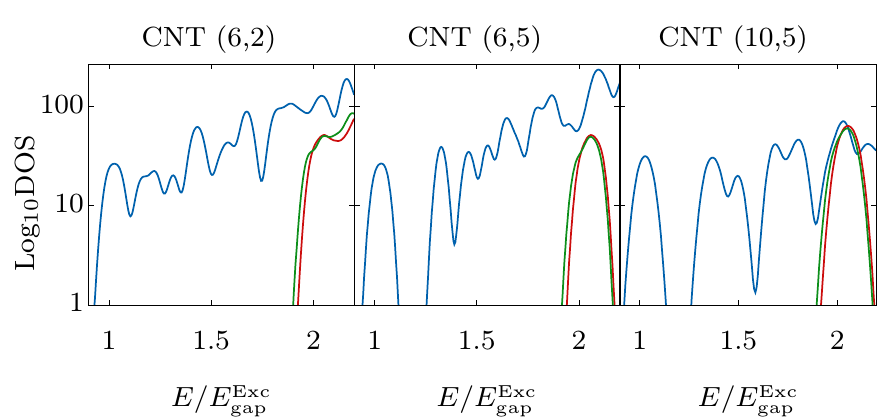} 
  \caption{ Exciton and biexciton DOS for the three nanotubes in units of exciton gap $E_{\rm gap}^{\rm exc}$ . Blue line is the exciton DOS, red line - biexciton DOS without self-energy corrections, green line - biexciton DOS with self-energy corrections. } \label{plots1}  
\end{figure}

Table \ref{table1} shows the results for the three SWCNTs. In all cases the resulting $\Sigma_{\alpha\beta}({E}_{\alpha} + {E}_{\beta})$ come out to be real, with vanishingly small imaginary parts. This is as expected since in our approximation we only include the possibility of the biexciton state decaying into an exciton and an electron-hole pair, which is suppressed due to energy conservation.     
The biexciton self-energy correction to the biexciton gap was obtained by evaluating Eqs. \eqref{expression} and \eqref{expression2} for a biexciton state made of two lowest-energy excitons, \textit{i.e.}, 
$\hbar\Sigma(E = 2 E_{\rm gap}^{\rm exc}),~E_{\rm gap}^{\rm exc}= {E}_1$.  For reference, the quasiparticle gap $E_{\rm gap}=\varepsilon_{LU}-\varepsilon_{HO}$ obtained from the DFT simulation is included. Figure \ref{plots1} shows plots of the density of states (DOS) for the excitons and biexcitons, including the shift due to exciton-exciton interactions. {The DOS was calculated using DOS$ = \sum_i \delta( E - E_i)$, where the delta function was approximated by a Gaussian function with a width corresponding to room temperature. For the exciton DOS, $E_i$ are the solutions to the Bethe-Salpeter equation - $E_{\alpha}$ (Eq. \eqref{BSE}). For biexciton DOS, $E_i = E_{\alpha} + E_{\beta}$ for a pair of non-interacting excitons and $E_i = E_{\alpha\beta}$ (Eq. \eqref{self_energy_corr}) for the self-energy corrected pair of excitons.  }
\begin{table}[H]
  \centering
  {\begin{tabular}{|c|c|c|c| } \hline
 Chirality & \textbf{$E_{gap}$} & $E_{\rm gap}^{\rm exc}$  & $\Sigma(2 E_{\rm gap}^{\rm exc})$   \\ \hline
$(6,2)$ & 1.33 & 0.98  & -0.045  \\ [1ex] \hline
$(6,5)$ & 1.22 & 1.09  & -0.041  \\ [1ex] \hline
$(10,5)$ & 0.91 & 0.835 & -0.036  \\ [1ex]\hline 
  \end{tabular}}
   \caption{ $E_{gap}$ is the HO-LU gap and $E_{\rm gap}^{\rm exc}=\epsilon_1$ is the lowest exciton energy obtained from BSE. $\Sigma(2 E_{\rm gap}^{\rm exc})$ is the biexciton self-energy for the lowest-energy biexciton.  All energies are in $eV$.} 
  \label{table1}
\end{table}

\noindent
Results show a small redshift in the biexciton density of states for all the nanostructures under consideration. As a percentage of the non-interacting gap, these shifts are: -4.59\% for (6,2),  -4.47\% for (6,5) and -4.31\% for (10,5) SWCNT. This is as expected since dipole-dipole electrostatic interactions are attractive.

In our approach it has been possible to sum perturbative contributions from the attractive interactions between the excitons in the biexciton state. However, the repulsive 
contributions are only included to the leading order. A calculation where both types of interactions are included to the leading order, {\it i.e.},  only including the contributions from 
the four diagrams in Fig. \ref{BIEXCfig}, has been performed. In this case the corrections to the biexciton gap as a fraction of 
the non-interacting gap are:  -1.84\% for (6,2),  -1.93\% for (6,5) and -2.27\% for (10,5) SWCNT. This suggests that better treatment of repulsive interactions by including 
higher order perturbative corrections, which is prohibitively expensive, would only reduce the negative shift in the biexciton gap.

\section{Conclusions and Outlook} \label{Conclusion}  
We have developed a first-principles DFT-based MBPT method to compute biexciton state energies in a semiconductor nanostructure. 
For that, we have computed self-energy of the biexciton state including (residual) electrostatic exciton-exciton interactions. In 
this work we have only included spin-zero excitons. These biexciton energies are relevant for, {\it e.g.}, accurate determination 
of the MEG threshold in a nanostructure.

To the first order in the Coulomb interaction there are four distinct Feynman diagrams contributing to the biexciton self-energy shown in Fig. \ref{BIEXCfig}. 
However, it has been possible to perform partial resummation of the perturbative corrections, such as those included in the exciton-electron coupling term 
in Eq. \ref{EXCFERMSF}. As a result, to the leading order in the electron-exciton coupling, just two distinct contributions that include intermediate exciton 
state appear, as shown Fig. \ref{BIEXCfig2}. Additionally, there are repulsive interactions between the like charges in the two excitons (electron-electron and hole-hole), 
which have been included to the leading order resulting in eight distinct contributions.
 
Calculations have been performed for the chiral SWCNTs (6,2), (6,5) and (10,5). 
We have found small negative corrections to the biexciton state gaps: -0.045 $eV$ in (6,2), which is 4.59\% of the non-interacting biexciton gap; -0.041 $eV$ in 
(6,5), which is 4.47\% of the non-interacting gap and -0.039 $eV$ in (10,5),  which is 4.31\%.
Small magnitude of energy shifts confirms validity of the perturbative approach for the biexciton states, at least for the SWCNTs,
and justifies simplified treatment of the biexcitons in SWCNTs as a pair of non-interacting excitons.

An important extension of the approach left to future work is to include the effects of triplet excitons, which would allow to compute energy of a biexciton 
made of a pair of a triplets in the overall singlet state. This is needed for precise determination of the MEG threshold in the SF channel. 

Another extension of this work would be to decorate both pairs of electron-hole lines in the diagrams in Fig. \ref{BIEXCfig2} with interactions (zigzag lines) and 
perform resummation. This would result in the processes where two intermediate exciton states appear. But small magnitude of the first-order biexciton 
energy corrections suggests that this modification of the technique is not likely to change the results significantly, at least for the SWCNTs.   

Improving the overall precision of the calculations preformed here can be done by the use of $G_0W_0$ calculations for single particle energies instead of using the HSE06 functional. 
This is expected to introduce an overall blueshift in both exciton and biexciton DOS, but not to alter our overall qualitative conclusions. Another step would be 
inclusion of the full, dynamically screened polarization function $\Pi(\omega,\kk,\pp)$ instead of the static, diagonal approximation $\Pi(0,-\kk,\kk)$. This would greatly 
increase computational cost, but it is not expected to alter the conclusions reached in this work.

\section{Acknowledgements}       
  Authors acknowledge financial support from the NSF grant CHE-1413614. 
The authors acknowledge the use of computational resources at
the Center for Computationally Assisted Science and
Technology (CCAST) at North Dakota State University. The diagrams shown in Figs. 1-5 have been created with JaxoDraw \cite{Jaxodraw}.     

\bibliography{mybib}  
\end{document}